\documentclass[prd,amsfonts,twocolumn,nofootinbib,showpacs]{revtex4}
\RequirePackage{graphicx}
\usepackage{enumerate}
\usepackage{comment}
 
\begin{document}
\title{Flavor changing nucleon decay}

\author{Nobuhiro Maekawa}
\affiliation{Kobayashi Maskawa Institute, Nagoya University; \\
Department of Physics, Nagoya University, Nagoya 464-8602, Japan}
\author{Yu Muramatsu}
\affiliation{School of Physics, KIAS, Seoul 130-722, Korea}

\begin{abstract}
Recent discovery of neutrino large mixings implies the large mixings in the
diagonalizing matrices of $\bf\bar 5$ fields in $SU(5)$ grand unified theory (GUT),
while the diagonalizing matrices of $\bf 10$ fields of $SU(5)$ are expected
to have small mixings like
Cabibbo-Kobayashi-Maskawa matrix. 
 We calculate the predictions of flavor changing nucleon decays (FCND) in $SU(5)$,
 $SO(10)$, and $E_6$ GUT models which have the above features for mixings. 
 We found that FCND can be the main
 decay mode and play an important role to test GUT models. 
\end{abstract}

\maketitle

\section{Introduction}
One of the most exciting discoveries in elementary particle physics among the latest 20 years 
is neutrino oscillation\cite{neutrino, Ue3}, which leads to
massive neutrinos and large neutrino mixing angles. Interestingly, this discovery gives an evidence of 
grand unified theory (GUT)\cite{GUT}, 
in which unification of forces and unification of quarks and leptons are
realized. This evidence for unification of quarks and leptons makes the idea of GUT quite promising,
because for unification of forces we have already known an experimental evidence that three gauge 
couplings meet at a scale, the GUT scale $\Lambda_G$\cite{unification},
especially in supersymmetric (SUSY) GUT\cite{SUSYGUT}.
 Moreover, the large neutrino mixing angles
imply not only large mixing angles of doublet lepton $l$  but also those of right-handed down quark 
$d_R^c$ in $SU(5)$ GUT because $\bf\bar 5$ field of $SU(5)$ contains $l$ and $d_R^c$.  This suggests
an interesting possibility that the flavor changing processes are seen in nucleon decay which is the
most important prediction of GUT
\footnote{
In Refs. \cite{Achiman}, in explicit GUT models which realize large
 neutrino mixings, the importance of $P\rightarrow\pi^0\mu^+$ has been
 discussed. Actually, in some models in Refs. \cite{Achiman}, $\Gamma(P\rightarrow\pi^0\mu^+)$ can be comparable to $\Gamma(P\rightarrow\pi^0e^+)$.}.
In this paper, we study the flavor changing nucleon decay and propose that the flavor changing nucleon decay can be a key observation for GUT.

\section{Qualitative evidence for $SU(5)$ unification}
First, we explain the qualitative evidence for the unification of matters. 
In $SU(5)$ GUT, Yukawa interactions with Higgs fields ${\bf 5}_H$ and
${\bf\bar 5}_H$ are given by
\begin{eqnarray}
{\cal L}_Y&=&(Y_u)_{ij} {\bf 10}_i{\bf 10}_j{\bf 5}_H+(Y_{d,e})_{ij}{\bf 10}_i{\bf \bar 5}_j{\bf\bar 5}_H
\nonumber \\
&+& (Y_{\nu_D}M_{\nu_R}^{-1}Y_{\nu_D}^t)_{ij}{\bf\bar 5}_i{\bf\bar 5}_j{\bf 5}_H{\bf 5}_H,
\end{eqnarray}
where $\bf 10$ fields contain doublet quark $q$, right-handed up quark
$u_R^c$, and right-handed charged lepton $e_R^c$, and the last term 
can be obtained from 
$(Y_{\nu_D})_{ij}{\bf \bar 5}_i{\bf 1}_j{\bf 5}_H+(M_{\nu_R})_{ij}{\bf 1}_i{\bf 1}_j$
by integrating the right-handed neutrino fields ${\bf 1}_i$. Here $i=1,2,3$ is the index for generation,
and $Y_u$, $Y_{d,e}$, $Y_{\nu_D}$, and $M_{\nu_R}$ are Yukawa matrix of up type quarks,  
that of down
type quarks and charged leptons, Dirac neutrino Yukawa matrix, and right-handed neutrino mass 
matrix, respectively.
These unified structures for Yukawa interactions are corresponding to the classification 
of the hierarchies of the observed quark and lepton masses that up type quark masses have the strongest
 hierarchy, neutrino masses have the weakest, and down type quark and charged lepton masses 
 have middle hierarchies if neutrino mass hierarchy is normal (not inverted). 
Moreover, if we assume that $\bf 10$
fields  induce stronger hierarchy in Yukawa couplings than $\bf\bar 5$ fields, these various 
hierarchies for quark and lepton masses can be explained. Furthermore, 
this assumption explains that quark mixings are smaller than lepton mixings at the same time 
if we use a
reasonable expectation that the stronger hierarchy leads to smaller mixings.
This brilliant chemistry between the Yukawa structure in $SU(5)$ GUT and the observed hierarchies
of quark and lepton masses and mixings is quite non-trivial, and therefore 
it can be regarded as an experimental
signature for unification of quarks and leptons in $SU(5)$ GUT.

\section{$E_6$ unification}
$E_6$ GUT\cite{E6,E6r,E6real} is more attractive because the assumption in the $SU(5)$ GUT can be derived, and
as a result, various Yukawa matrices can be derived from one basic Yukawa hierarchy\cite{E6real}.
The
fundamental representation in $E_6$ is divided into $SO(10) (SU(5))$ representations as
\begin{equation}
{\bf 27}={\bf 16}({\bf 10}+{\bf\bar 5}+{\bf 1})+{\bf 10}({\bf 5}+{\bf\bar 5}')+{\bf 1}({\bf 1}).
\end{equation}
This ${\bf 27}$ includes one generation quarks and leptons in addition to one pair of vector-like 
fields ${\bf 5}+{\bf\bar 5}$ and a singlet.
If we introduce three ${\bf 27}_i$ ($i=1,2,3$) for three generation quarks and leptons, we have
six $\bf\bar 5$ fields. Three of six $\bf\bar 5$ fields become superheavy after 
developing the
vacuum expectation values (VEVs) of ${\bf 27}_H$ and ${\bf 27}_C$ through the Yukawa interactions
\begin{equation}
W_Y=(Y^H)_{ij}{\bf 27}_i{\bf 27}_j{\bf 27}_H+(Y^C)_{ij}{\bf 27}_i{\bf 27}_j{\bf 27}_C,
\end{equation}
where the VEV of ${\bf  27}_H$ breaks $E_6$ into $SO(10)$ and the VEV of  ${\bf 27}_C$ breaks 
$SO(10)$ into $SU(5)$. Once we fix $Y^H$, $Y^C$,  $\langle {\bf 27}_H\rangle$, and 
$\langle {\bf 27}_C\rangle$, $3\times 6$ mass matrix of three $\bf 5$s and six $\bf\bar 5$s is
determined, and therefore, three massless modes ${\bf\bar 5}_i^0$ are fixed.
Here we assume that these Yukawa
couplings $Y^H$ and $Y^C$ have strong hierarchy corresponding to the hierarchy of ${\bf 10}$ of
$SU(5)$. Typically, we take
\begin{equation}
Y^H\sim Y^C\sim \left(\begin{array}{ccc} \lambda^6 & \lambda^5 & \lambda^3 \\
                                                       \lambda^5 & \lambda^4 & \lambda^2 \\
                                                       \lambda^3 & \lambda^2 & 1 
                                \end{array}\right),
\end{equation}
where a unit of hierarchy $\lambda\sim 0.22$ is taken to be around the Cabibbo mixing to obtain Cabibbo-Kobayashi-Maskawa (CKM) matrix\cite{CKM}.
The $O(1)$ coefficients are
omitted usually in this paper. 
Then, two ${\bf\bar 5}$ fields from ${\bf 27}_3$ become superheavy unless
$\langle {\bf 27}_H\rangle \ll \langle {\bf 27}_C\rangle$ because they have larger Yukawa 
couplings and therefore have larger mass parameters.
The three massless modes ${{\bf\bar 5}_i^0}$ come from the first two generation fields ${\bf 27}_1$
and ${\bf 27}_2$ which have smaller Yukawa couplings.
As a result, three ${{\bf\bar 5}_i^0}$, whose main modes typically become 
$({\bf\bar 5}_1, {\bf\bar 5'}_1, {\bf\bar 5}_2)$, induce milder Yukawa hierarchy than ${{\bf 10}_i}$ fields,
that is nothing but what we assume in the $SU(5)$ GUT to obtain realistic hierarchies of quark and lepton masses and mixings. 
Note that ${{\bf\bar 5}_2^0}\sim {{\bf\bar 5'}_1}+\lambda^\Delta{{\bf\bar 5}_3}$ has Yukawa couplings 
through the mixing with ${\bf\bar 5}_3$ when the Higgs ${\bf 5}_H$ and ${\bf\bar 5}_H$ are included
in ${\bf 10}_H$ of $SO(10)$ in ${\bf 27}_H$. Then we can obtain realistic Yukawa hierarchies
as
\begin{equation}
Y_u\sim \left(\begin{array}{ccc} \lambda^6 & \lambda^5 & \lambda^3 \\
                                                       \lambda^5 & \lambda^4 & \lambda^2 \\
                                                       \lambda^3 & \lambda^2 & 1 
                                \end{array}\right),
Y_d\sim Y_e^t\sim Y_{\nu_D}^t\sim \left(\begin{array}{ccc}
 \lambda^6 & \lambda^{\Delta+3} & \lambda^5 \\
 \lambda^5 & \lambda^{\Delta+2} & \lambda^4 \\
 \lambda^3 & \lambda^\Delta & \lambda^2 
 \end{array}\right),
\end{equation}
when $\Delta\sim 2.5$.
The right-handed neutrino masses are obtained from
\begin{equation}
\frac{(Y^{XY})_{ij}}{\Lambda}{\bf 27_i}{\bf 27_j}{\bf\overline{27}}_{X}{\bf\overline{27}}_{Y},
\end{equation}
where $X,Y=\bar H, \bar C$, $\Lambda$ is the cutoff scale, after developing the
VEVs $|\langle {\bf 27}_H\rangle|=|\langle {\bf\overline{27}}_{\bar H}\rangle|$ and
$|\langle {\bf 27}_C\rangle|=|\langle {\bf\overline{27}}_{\bar C}\rangle|$. 
Here we take $Y^{XY}\sim Y^H\sim Y^C$. 
All quark and lepton mass matrices can be diagonalized by unitary matrices
for ${\bf 10}$ fields and ${\bf\bar 5}$ fields
\begin{equation}
V_{\bf 10}\sim \left(\begin{array}{ccc} 1 & \lambda & \lambda^3 \\
                                                       \lambda &1 & \lambda^2 \\
                                                       \lambda^3 & \lambda^2 & 1 
                                \end{array}\right), 
 V_{\bf\bar 5}\sim \left(\begin{array}{ccc} 1 & \lambda^{3-\Delta} & \lambda \\
                                                       \lambda^{3-\Delta} &1 & \lambda^{\Delta-2} \\
                                                       \lambda & \lambda^{\Delta-2} & 1 
                                \end{array}\right),
 \end{equation}
and we can obtain realistic CKM matrix $V_{\rm CKM}\sim V_{\bf 10}$ and 
the Maki-Nakagawa-Sakata (MNS) matrix\cite{MNS}
 $V_{\rm MNS}\sim V_{\bf\bar 5}$,
 when $\Delta\sim 2.5$.  Note that the important prediction 
 $(V_{\rm MNS})_{13}\sim (V_{\rm CKM})_{12}$, 
 which was confirmed
 by recent neutrino experiments as 
 $(V_{\rm MNS})_{13}\sim 0.15$\cite{Ue3}, is caused by 
 ${{\bf\bar 5}_3^0}\sim {{\bf\bar 5}_2}$. Therefore, to obtain the realistic hierarchies of quark and lepton
 masses and mixings, it is essential that the ${\bf\bar 5'}_1$, which comes from $\bf 10$ of $SO(10)$,
 becomes the second generation $\bf\bar 5$ field ${\bf\bar 5}_2^0$. That structure is important to study of the prediction of the 
 nucleon decay in the next section.

Note that the relation ${\bf\bar 5}_2^0\sim {\bf\bar 5'}+\lambda^\Delta{\bf\bar 5}_3$ can be realized 
even in $SO(10)$ unification,  if  
$\bf 10$ of $SO(10)$, which provides $\bf\bar 5'$,  is introduced as a matter field\cite{SO10}.
Therefore, we have three GUT models which satisfy the Yukawa hierarchy
hypothesis,
 ``${\bf 10}$ fields induce stronger hierarchy in Yukawa couplings than $\bf\bar 5$ fields''. 
 Their unification groups are
$SU(5)$, $SO(10)$, and $E_6$. Next, we study how to identify these
unification group by observing various partial nucleon decay widths.

\section{Nucleon decay}
 In this paper, we concentrate on the nucleon decay via dimension 6 operators\cite{dim6ope}, because
 the nucleon decay via dimension 5 operators\cite{dim5} is strongly dependent on the explicit model
 of GUT Higgs sector which is expected to have big modification to solve the most difficult problem 
 called the doublet-triplet splitting problem\cite{DTsplitting} and 
 because it is strongly suppressed in natural GUT
 in which the difficult problem is solved with natural 
assumption\cite{SO10, GCU, naturalGUT}.
 
The dimension 6 effective operators which induce nucleon decay in $E_6$ GUT are produced via mediation by
$SU(5)$ superheavy gauge boson $X$, $SO(10)$ superheavy gauge boson $X'$, and $E_6$ 
superheavy gauge boson $X''$ as\cite{MM}
\begin{eqnarray}
\mathcal{L}_{eff}&=&
\frac{g_{G}^2}{M_X^2}\left\{ \right.
-(\overline{e^c_R}_i u_{Rj})(\overline{q^c}_j q_{i})+
(\overline{l^c}_i q_{j})(\overline{u_R^c}_j d_{Ri}) \nonumber\\&+&
(\overline{L^c}_i q_{j})(\overline{u_R^c}_j D_{Ri}) \left. \right\}
+\frac{g_{G}^2}{M_{X'}^2}
(\overline{l^c}_i q_{j})(\overline{u_R^c}_i d_{Rj})
\nonumber\\ 
&+&\frac{g_{G}^2}{M_{X''}^2}
(\overline{L^c}_i q_{j})(\overline{u_R^c}_i D_{Rj})
\label{dim6}
\end{eqnarray}
where $g_{G}$ is the unified gauge coupling and 
the superheavy gauge boson masses $M_X$, $M_{X'}$, and $M_{X''}$ are dependent on the 
VEVs of the GUT Higgs which break $E_6$ into the SM gauge group. 
Here, large character denotes $\bf\bar 5'$ field which comes from $\bf 10$ of $SO(10)$.
In the $SO(10)$ GUT, we just take $M_{X''}\rightarrow \infty$, and in $SU(5)$ GUT, we take $M_{X'}, M_{X''}\rightarrow \infty$ and neglect the 
interactions which include the large character fields.
Note that the nucleon decay via dimension 6 operators depends on Yukawa couplings, although
this is via gauge interactions. The situation is similar to the weak interaction. The weak interaction
is also the gauge interaction, but we have CKM mixings which are determined by Yukawa couplings. 
For the nucleon decay, the nucleon decay via dimension 6 operators depends on the diagonalizing
matrices for Yukawa matrices. 
However, we have already understood the mixings in GUT as the qualitative evidence for 
the $SU(5)$ GUT. Especially for the diagonalizing matrices, $V_{\bf10}$ and $V_{\bf\bar 5}$ are
fixed as CKM matrix and MNS matrix, respectively, except $O(1)$ coefficients.
Therefore, these ambiguities are almost fixed by our
understanding of Yukawa structures. 
Therefore, 
we can compare the predictions of nucleon decays in $SU(5)$, $SO(10)$ and $E_6$ GUTs.

Important observation to find useful nucleon decay modes for identification of unification group
is that all four fermions in the first term in Eq. (\ref{dim6}) 
come from $\bf 10$ of $SU(5)$ fields, and in the other terms 
 two of four fermions come from $\bf\bar 5$ fields. Since $X'$ and $X''$ 
gauge interactions induce only the effective interactions with $\bf\bar 5$ 
fields,  we should look for the nucleon decay
 modes in which the operators with ${\bf\bar 5}$ fields are significant to 
 identify the unification group.
 
\begin{figure}[t]
\centering
\includegraphics[width=1.0\columnwidth]{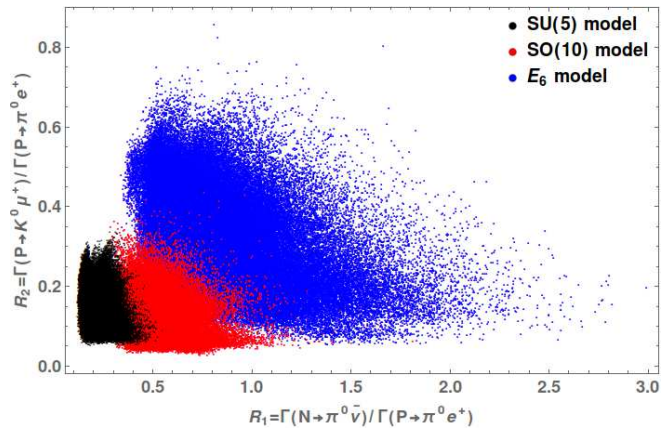}
 \caption{The distribution of $10^5$ model points for $SU(5)$(black), $SO(10)$(red), and $E_6$(blue) GUTs with
 horizontal axis $R_1=\Gamma(N\rightarrow \pi^0\bar\nu)/\Gamma(P\rightarrow \pi^0e^+)$ and 
 vertical axis 
 $R_2=\Gamma(P\rightarrow K^0\mu^+)/\Gamma(P\rightarrow \pi^0e^+)$. The superheavy gauge
 boson masses are taken to be $M_X=M_{X'}=\sqrt{2}M_{X''}$. }
\label{fig:R1R2}
\end{figure}

Since all operators with ${\bf\bar 5}$ fields include a lepton doublet while ${\bf 10}$ field includes
no neutrino, the modes with neutrino can be important to identify the unification group. 
The decay mode $N\rightarrow \pi^0 \bar \nu$
\footnote{In the decay modes which include neutrino, 
we sum up over the flavor of neutrino because 
the nucleon decay detectors do not distinguish neutrino types.}
 has been studied in the literature for the
 identification\cite{Langacker, MM}. Especially in Ref\cite{MM}, we have shown that two ratios 
 $R_1\equiv\Gamma(N\rightarrow \pi^0\bar\nu)/\Gamma(P\rightarrow \pi^0e^+)$
and $R_2\equiv\Gamma(P\rightarrow K^0\mu^+)/\Gamma(P\rightarrow \pi^0e^+)$ are useful to identify
three unification group as in Fig. \ref{fig:R1R2}, where we have $10^5$ model points for each unification
group $SU(5)$(black points), $SO(10)$(red points),
and $E_6$(blue points) and the magnitudes of the $O(1)$ coefficients of diagonalizing matrices 
are determined randomly
between 0.5 and 2.  We adopt superheavy gauge boson masses $M_X=M_{X'}=\sqrt{2}M_{X''}$ as in the previous paper\cite{MM}.
\footnote{
In this paper, we have not fixed $V_{u_R^c}=1$ 
(and $V_{d_R^c}=1$ for $SU(5)$), which are adopted in Ref. \cite{MM}.
Theoretically we can fix those diagonalizing
matrices without loss of generality. If we have not imposed any constraints
to the other diagonalizing matrices, it would not produce any changes in
the results. However, in our analysis, we constrained the $O(1)$ 
coefficients of the other diagonalizing matrices, and therefore, the results
depends on whether these conditions are imposed or not.
We think that the results without these conditions become similar to
the results with these conditions with wider allowed range for the 
$O(1)$ coefficients. Therefore, distributions of model points have 
become wider in this paper than in the previous one.
}  
In the calculations in this paper,
we use the hadron matrix elements calculated
by lattice\cite{lattice}, and the renormalization factors of the minimal SUSY
$SU(5)$ GUT as $A_R=3.6$ for the operators which include a right-handed
charged lepton $e_R^c$ and $A_R=3.4$ for the operators which include the 
doublet leptons $l$ as the reference values\cite{RF}.
The ratio $R_2$ is sensitive to flavor structure of the second generation,
and very useful to identify $SO(10)$ and $E_6$ unification group.
Interestingly, $R_1$ can be larger than one especially for higher rank unification group like
$E_6$. Of course the results are strongly dependent on the mass spectrum of superheavy gauge
bosons. If $M_{X''}\gg M_{X'}=M_X$, the $E_6$ model points shrink to $SO(10)$ model points, and when $M_{X'}$ becomes much larger than $M_X$, the $SO(10)$ model points shrink to
the $SU(5)$ model points. However, we can say that if $R_1>0.5$, $SU(5)$ is implausible and if
$R_1>1$, $E_6$ is preferable.
Unfortunately, the detection efficiency for the mode $N\rightarrow \pi^0\bar\nu$
is not so high as $P\rightarrow \pi^0e^+$ mode\cite{SKppie,SKnpinu}
, and therefore, it requires
extremely more powerful experiments to
observe the mode $N\rightarrow \pi^0\bar\nu$ even if $R_1>1$.

\begin{figure}[t]
\centering
\includegraphics[width=1.0\columnwidth]{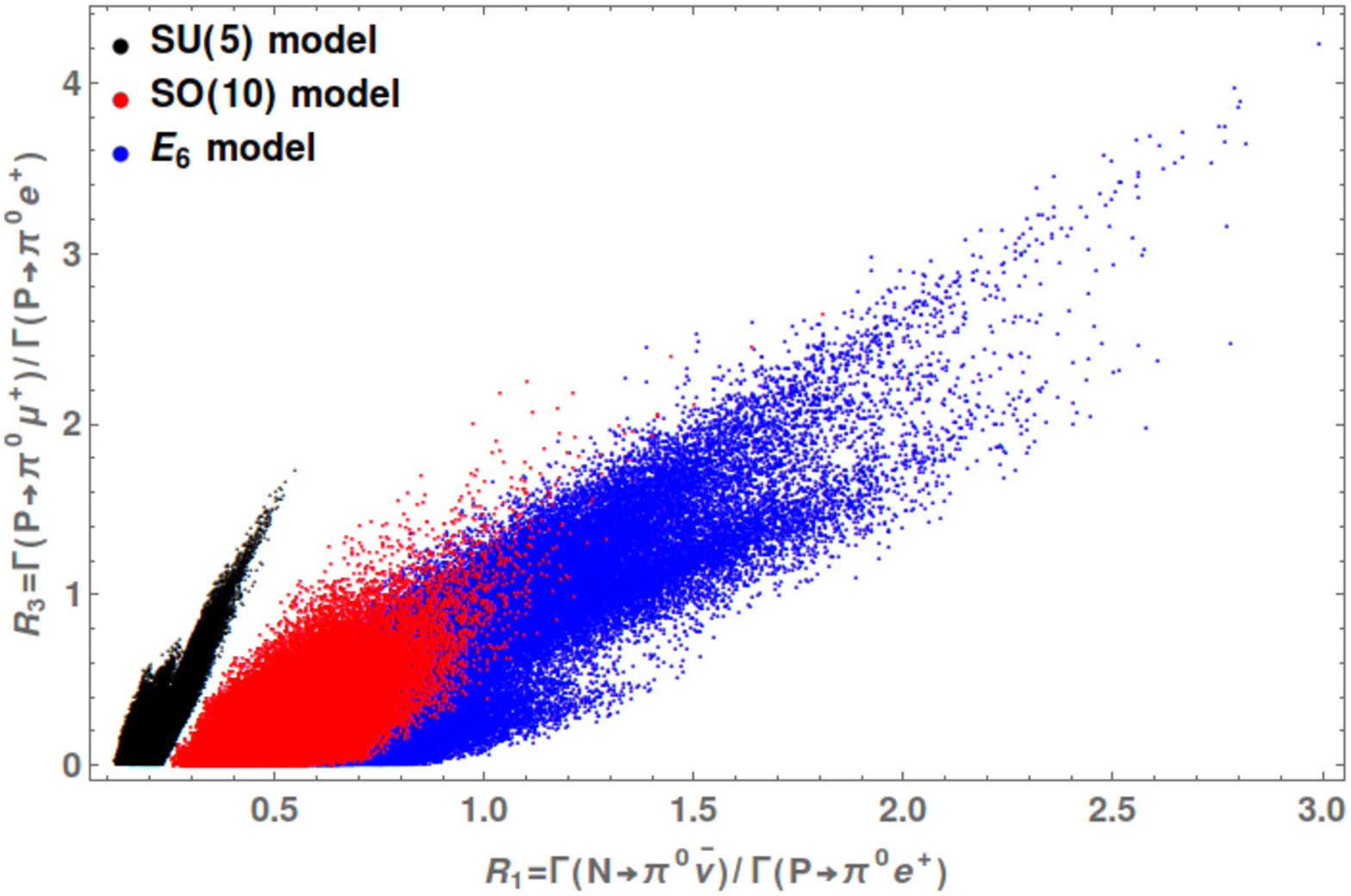}
 \caption{The distribution of $10^5$ model points for $SU(5)$(black), $SO(10)$(red), and $E_6$(blue) GUTs with
 horizontal axis $R_1=\Gamma(N\rightarrow \pi^0\bar\nu)/\Gamma(P\rightarrow \pi^0e^+)$ and 
 vertical axis 
 $R_3=\Gamma(P\rightarrow \pi^0\mu^+)/\Gamma(P\rightarrow \pi^0e^+)$. The superheavy gauge
 boson masses are taken to be $M_X=M_{X'}=\sqrt{2}M_{X''}$. }
\label{fig:R1R3}
\end{figure}

\begin{figure}[t]
\centering
\includegraphics[width=1.0\columnwidth]{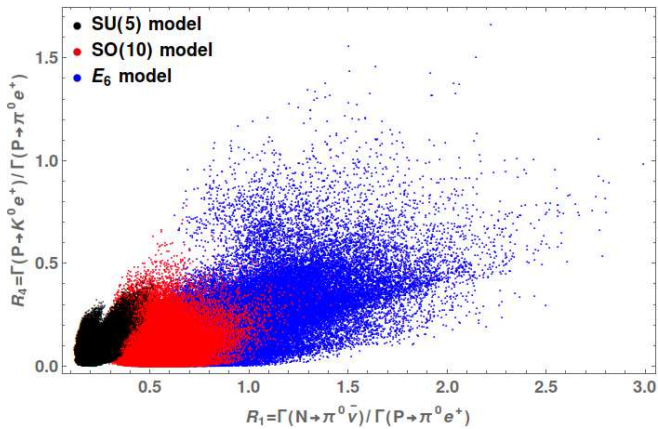}
 \caption{The distribution of $10^5$ model points for $SU(5)$(black), $SO(10)$(red), and $E_6$(blue) GUTs with
 horizontal axis $R_1=\Gamma(N\rightarrow \pi^0\bar\nu)/\Gamma(P\rightarrow \pi^0e^+)$
  and vertical axis $R_4\equiv\Gamma(P\rightarrow K^0e^+)/\Gamma(P\rightarrow \pi^0e^+)$
  The superheavy gauge
 boson masses are taken to be $M_X=M_{X'}=\sqrt{2}M_{X''}$. }
\label{fig:R1R4}
\end{figure}

In this paper, we propose novel modes which may be useful for the identification of unification group. 
Essential point is that $\bf\bar 5$ fields have large mixings in diagonalizing matrices while
$\bf 10$ fields have small mixings. And therefore, flavor changing nucleon decay, for example,
$P\rightarrow \pi^0\mu^+$ or $P\rightarrow K^0 e^+$, becomes
more important for higher rank unification group. 
In Figs. \ref{fig:R1R3} and \ref{fig:R1R4},
we have calculated the two ratios 
$R_3\equiv\Gamma(P\rightarrow \pi^0\mu^+)/\Gamma(P\rightarrow \pi^0e^+)$
and $R_4\equiv\Gamma(P\rightarrow K^0e^+)/\Gamma(P\rightarrow \pi^0e^+)$ 
with horizontal axis $R_1$ in $10^5$ model points of $SU(5)$(black), 
$SO(10)$(red), and $E_6$(blue) GUTs with 
the superheavy gauge boson masses $M_{X}=M_{X'}=\sqrt{2}M_{X''}$.
Interestingly, the $SU(5)$ model points are clearly separated from
$SO(10)$ and $E_6$ model points in Fig. \ref{fig:R1R3}, while Fig. \ref{fig:R1R2} has no such separation.
One more interesting point is that there are a lot of model points with $R_3>1$.
Since the detection efficiency of the $P\rightarrow \pi^0\mu^+$ is as large as that of 
$P\rightarrow\pi^0e^+$\cite{SKppie},  the flavor changing nucleon decay mode 
$P\rightarrow \pi^0\mu^+$ can be found earlier than
$P\rightarrow \pi^0e^+$ if $R_3>1$.
On the contrary, $R_4$ is comparatively smaller, mainly because the mode
$\Gamma(P\rightarrow K^0e^+)$ has the phase space suppression and
smaller hadron matrix elements.
Note that there is a tendency to obtain larger $R_1$ for larger $R_3$.

Although it may not be so clear in these figures, GUT with larger rank unification group predicts larger FCND.
Actually, it is seen in concrete numbers of model points with $R_3>1$ (17\% in $E_6$, 0.7\% in $SO(10)$ and
0.5\% in $SU(5)$).

It must be useful to stress the advantage of the neutrino modes like
$N\rightarrow \pi^0\bar\nu$ for identification of unification group,
although such modes have disadvantage for the detection.
The most important feature for $\Gamma(N\rightarrow \pi^0\bar\nu)$
is that the value becomes larger for GUT with larger rank unification
group, especially when $\bf 10$ fields have small mixings. Actually,
when $V_{\bf 10}=1$, we can show that
\begin{equation}
\frac{\Gamma(N\rightarrow \pi^0\bar\nu)}
{\Gamma_{SU(5)}(N\rightarrow \pi^0\bar\nu)}=1
+\alpha(2+\alpha)|(V_{d_R^c})_{11}|^2
+\beta(2+\beta)|(V_{d_R^c})_{21}|^2,
\end{equation}
where $\alpha\equiv M_{X}^2/M_{X'}^2$ and $\beta\equiv M_X^2/M_{X''}^2$.
Here we have summed the flavor of neutrinos, that is important in this 
calculation.
Obviously $R_1$ becomes larger for larger unification group.
This feature is quite important to identify the unification group.

\section{Discussion and summary}
\begin{figure}[t]
\centering
\includegraphics[width=1.0\columnwidth]{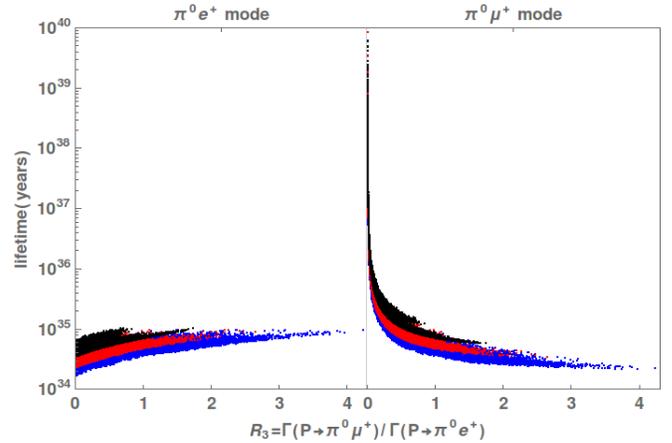}
 \caption{The distribution of $10^5$ model points for $SU(5)$ (black), 
 $SO(10)$ (red), and $E_6$ (blue) GUTs with 
 $M_X/g_{\rm G}=1\times 10^{16}$ GeV.
 Horizontal axis is $R_3=\Gamma(P\rightarrow \pi^0\mu^+)/\Gamma(P\rightarrow \pi^0e^+)$
  and vertical axis is partial lifetime for $P\rightarrow \pi^0e^+$ and $P\rightarrow \pi^0\mu^+$,
  which are proportional to $(M_X/g_{\rm G})^4$. 
  The superheavy gauge
 boson masses are taken to be $M_X=M_{X'}=\sqrt{2}M_{X''}$. }
\label{fig:lifetime}
\end{figure}

Recently, two events have been found in the signal region for the process 
$P\rightarrow \pi^0\mu^+$\cite{ND}, 
though these are still consistent with the background  expected to be 0.9 event 
mainly from atmospheric neutrino events.
If the signature for the flavor changing nucleon decay $P\rightarrow \pi^0\mu^+$ 
has been found in SuperKamiokande, higher rank unification group like $SO(10)$ or
$E_6$ is preferable 
when the mixings of $\bf 10$ fields are small.
The predicted partial lifetime for $M_X/g_{\rm G}=1\times 10^{16}$ GeV is presented
in Fig. \ref{fig:lifetime}.
Obviously, for larger $R_3$, longer partial lifetime for $P\rightarrow \pi^0e^+$ and shorter partial lifetime for $P\rightarrow \pi^0\mu^+$ are
obtained. In $SU(5)$, both partial lifetimes become longer than in $SO(10)$
and $E_6$. 
If the signature is from the real nucleon decay process, 
it is obvious that the usual MSSM predicted value 
$M_X/g_{\rm G}\sim 3\times 10^{16}$ is too large to explain the events
even if the ambiguities in Hadron matrix elements\cite{lattice} are taken 
into account.
Therefore, to explain the signal, larger unification gauge coupling $g_X$
(it requires extra vector-like fields in addition to the MSSM fields.), and/or
smaller superheavy gauge boson mass $M_X$ are required.
Note that both features are predicted in the natural GUT
\cite{SO10, GCU, naturalGUT}, in which the nucleon decay via dimension
6 operators is enhanced while that via dimension 5 is suppressed.

Which mode will be found next?
We expect that $\Gamma(N\rightarrow \pi^0\bar\nu)$ can be larger than 
$\Gamma(P\rightarrow \pi^0e^+)$, 
since $R_3$ is positively correlated with $R_1$ as in Fig.  \ref{fig:R1R3}.
However, since the detection efficiency for the mode $N\rightarrow \pi^0\bar\nu$ is much smaller
than that for $P\rightarrow \pi^0e^+$, we can predict that next mode should be 
$P\rightarrow\pi^0e^+$.
Of course, the other modes, $N\rightarrow \pi^0\bar\nu$ and $P\rightarrow K^0e^+$,
are expected to be found in future experiments like HyperKamiokande \cite{HK}.
The observation of these modes is quite important and gives us critical hints
for studying GUT models.

In this paper, we have emphasized the importance of flavor changing
nucleon decay, whose observation may identify the unification group.
Especially, the mode $P\rightarrow \pi^0\mu^+$ is important because
the detection efficiency is as large as the usual mode 
$P\rightarrow \pi^0e^+$. The partial lifetime of $P\rightarrow \pi^0\mu^+$
can be shorter than that of $P\rightarrow \pi^0e^+$ especially
in $E_6$ GUT.
Of course, our results are strongly dependent on our important 
assumptions for diagonalizing matrices of quarks and leptons,
$V_{\bf10}\sim V_{CKM}$ and $V_{\bf\bar 5}\sim V_{MNS}$, and
for the unification group which are restricted to $SU(5)$, $SO(10)$, and
$E_6$. Therefore, our results are not directly applied to the models which
do not satisfy our assumptions like the GUT models in Refs. \cite{Achiman}.

Although most of model points predict longer partial
lifetime of $P\rightarrow \pi^0\mu^+$ than that of $P\rightarrow \pi^0e^+$, it is important to pay attention to 
the mode $P\rightarrow \pi^0\mu^+$ even if the present signal for
$P\rightarrow \pi^0\mu^+$ is from the back ground processes.

\section{Ackowledgement}
This work is supported in part by 
National Research Foundation of Korea (NRF) Research Grant
NRF- 2015R1A2A1A05001869 and Grants-in-Aid for Scientific Research from 
MEXT of Japan(No. 15K05048).

\end{document}